\documentclass[prd,nofootinbib,english,superscriptaddress,10pt]{revtex4}%twocolumn
\usepackage{amsfonts,amsmath,hyperref,url, color}
\hypersetup{
colorlinks=true,
citecolor=black,
linkcolor=black,
urlcolor=black
}
\usepackage{bm, bbm}
\usepackage{graphicx}
\usepackage{mathtools}
\usepackage{nccmath}
\usepackage[font=small,labelfont=bf]{caption}
\usepackage{tabularx}

\usepackage[utf8]{inputenc}
\usepackage[english]{babel}
\usepackage{amsmath,amsfonts,amssymb}
\usepackage{hyperref}
\usepackage{caption,adjustbox}
\usepackage{multirow}
\usepackage{cleveref}
\usepackage{fontawesome}
\usepackage{comment}
\usepackage{graphicx}
\usepackage{float}
\usepackage{bbding}
\usepackage[usenames,dvipsnames]{xcolor}
\usepackage[arrowdel]{physics}
\usepackage{tensor}
\usepackage{cancel}
\usepackage{tikz}
\usepackage{cleveref}
\usepackage{mathrsfs}
\usepackage{listings} 
\usepackage[title,toc,titletoc]{appendix}
\usepackage{titlesec}
\usepackage{tocloft}
\usetikzlibrary{decorations.pathmorphing, patterns,shapes, decorations.pathreplacing}

	\usepackage{amstext,amsthm,verbatim,cancel,epsfig} 
\usepackage{color}

\newcommand{\be}{\begin{equation}}
\newcommand{\ben}{\begin{equation*}}
\newcommand{\ee}{\end{equation}}
\newcommand{\een}{\end{equation*}}
\newcommand{\ba}{\begin{eqnarray}}
\newcommand{\ea}{\end{eqnarray}}

\newcommand{\bal}{\begin{align}}
\newcommand{\eal}{\end{align}}

\newcommand{\bea}{\begin{eqnarray}}
\newcommand{\eea}{\end{eqnarray}}
\newcommand{\bit}{\begin{itemize}}
\newcommand{\eit}{\end{itemize}}
\newcommand{\la}{\langle}
\newcommand{\ra}{\rangle}

\graphicspath{ {./images} }

\begin{document}

\title{Quantum evolution of Hopf algebra Hamiltonians}

%\title{On the quantum evolution of Hopf algebra hamiltonians}

%\title{Do Hopf algebra hamiltonians describe a physically viable quantum evolution?}

\author{Michele Arzano}
\email{michele.arzano@na.infn.it}
\affiliation{Dipartimento di Fisica ``E. Pancini", Universit\`a di Napoli Federico II, I-80125 Napoli, Italy\\}

\affiliation{INFN, Sezione di Napoli, Complesso Universitario di Monte S. Angelo,
Via Cintia Edificio 6, 80126 Napoli, Italy}

\author{Antonio Del Prete}
\email{delpreteantonio1601@gmail.com}

\affiliation{Dipartimento di Fisica ``E. Pancini", Universit\`a di Napoli Federico II, I-80125 Napoli, Italy\\}

\author{Domenico Frattulillo}
\email{domenico.frattulillo@unina.it}

\affiliation{Dipartimento di Fisica ``E. Pancini", Universit\`a di Napoli Federico II, I-80125 Napoli, Italy\\}

\affiliation{INFN, Sezione di Napoli, Complesso Universitario di Monte S. Angelo,
Via Cintia Edificio 6, 80126 Napoli, Italy}

\begin{abstract}
In recent years, growing attention has been devoted to the possibility that theories with deformed symmetries, associated with certain models of non-commutative spacetime, may encode a fundamental form of decoherence. This effect should be described by a Lindblad-like evolution governed by the non-trivial Hopf algebra structure of the time-evolution generators. In this work we provide a detailed analysis of such possibility for similar Hopf algebra deformations of the Hamiltonian of a qubit. Starting from a critical examination of the very definition of time evolution through the generalized adjoint action, we explore whether a coherent and physically viable framework can be established. In particular, our analysis shows that a more general combination of adjoint actions always guarantees a von Neumann dynamics and, also in the case of deformed spacetime symmetries considered in the literature, a physically viable Lindblad evolution cannot be established.
\end{abstract}

\maketitle
\section{Introduction}
It is widely believed that, in a putative theory of quantum gravity, the familiar description of spacetime as a smooth manifold should inevitably break down, giving way to a more fundamental, possibly discrete or non-commutative, structure \cite{Snyder:1946qz, Doplicher:1994tu}.
Among the multitude of theories and effective models developed to describe physics at the Planck scale, frameworks based on deformed symmetries and non-commutative spacetime structures have acquired a prominent role, providing a concrete setting in which possible Planck scale modifications of spacetime can lead to experimentally testable effects \cite{Addazi:2021xuf,Arzano:2021scz}.

In the literature exploring the phenomenology of models of Planck-scale departures from ordinary Lorentz symmetry, most of the attention  has been devoted to modifications of the energy-momentum dispersion relation and time of arrivals of the highest energy particles of astrophysical origin \cite{Amelino-Camelia:2008aez} (see also \cite{Addazi:2021xuf} and references therein). In recent years, however, a growing number of studies have focused on potential effects of deformed symmetry frameworks involving departures from CPT symmetries and fundamental decoherence \cite{Arzano:2019toz,Arzano:2020rzu,Bevilacqua:2024jpy,Arzano:2022nlo}.
These effects are based on specific features of deformed symmetries, namely their non-trivial Hopf algebra structure. The connection between space-time non-commutativity and Hopf algebra deformations of symmetries has been widely explored, starting with the seminal work \cite{Majid:1994cy}, which introduced the noncommutative $\kappa$-Minkowski spacetime associated to the $\kappa$-Poincaré algebra \cite{Lukierski:1992dt} (an analogous connection with a Hopf algebra deformation of Poincaré symmetries was later established also for canonical $\theta$-Minkowski noncommutative spacetime \cite{Chaichian:2004za}).
We do not intend in the present work to review the mathematical details of this connection, but, as we will briefly review in the next section, Hopf algebra structures at a quantum mechanical level provide the prescriptions for defining the action of symmetry generators and observables on composite systems and on the dual space of the Hilbert space of states of a quantum system. 

In this spirit, in \cite{Arzano:2014cya} it was observed that modifications of the standard Hopf algebra structures also affect the adjoint action of symmetry generators on operators and, in particular, of generators of time evolution on the density matrix. Specifically, it was suggested that in the context of the $\kappa$-Poincaré algebra, for a specific choice of algebra generators known as the ``classical basis" \cite{Borowiec:2009vb}, generators of time translations act non-trivially on density matrices, leading to an adjoint action which is no longer a commutator but assumes a ``deformed" Lindblad‑like form.

More recently this claim was re-examined in \cite{Arzano:2022nlo}. In the work, it was observed that, taking an appropriate non-relativistic limit of the $\kappa$-Poincaré algebra and adopting a symmetrized version of the adjoint action to define the time evolution, one can obtain a genuine Lindblad evolution with a positive semi-definite Lindblad term. In this way, the model realizes a scenario in which pure states evolve irreversibly into mixed states, even in the absence of an external environment, offering a concrete example of how Planck-scale effects can leave an imprint on the general dynamics of quantum systems. Although in the work the authors obtain a formally valid Lindblad equation, they apply a map to the symmetry generators, which, as we will discuss in detail at the end of our analysis, leads to complex values for energy and momenta of composite systems, making the model physically inconsistent.

%In the cited references, particular prescriptions for the definition of evolution of the quantum state have been proposed. 

The aim of the present work is to provide a detailed analysis of the possibility of obtaining deformed time evolution equations (possibly of the Lindblad type) by studying the deformed Hopf algebra of observables of a much simpler system: {\it a qubit}. The deformed observables for such system are described by $q$-deformations of the $\mathfrak{su}(2)$ algebra. This choice is motivated by both mathematical and phenomenological considerations. On the one hand, the structural simplicity of such deformed algebra makes it an excellent testing ground for investigating how modified symmetries can induce departures from standard quantum time evolution. On the other hand, recent works have explored the physical implications of similar deformations of the $SU(2)$ group in scenarios beyond standard quantum mechanics \cite{Amelino-Camelia:2022dsj, DEsposito:2024wru,Arzano:2025zsp}.

For this purpose, after reviewing the notion of time evolution of quantum systems from an Hopf algebra point of view in the next section, in Section III we consider the most studied $q$-deformation of the universal enveloping algebra of the $\mathfrak{su}(2)$ Lie algebra \cite{biedenharn1995quantum}. We critically analyze the definition of time evolution through a generalized version of the adjoint action, assessing the possibility of obtaining a physically viable time evolution equation. Our results show that only a specific combination of left and right adjoint actions of the Hamiltonian and their hermitian adjoints leads to a physically viable evolution equation which is just the ordinary von Neumann one. 

In Section IV we construct the most general Hopf-algebra deformations of $su(2)$ with undeformed algebraic sector, at leading order in the deformation parameter, obtaining, again, the result that there is only a particular prescription which guarantees an admissible physical evolution for all the parameters. Interestingly, all the resulting time evolutions are not of the Lindblad type but coincide with the standard von Neumann equation governed by an undeformed Hamiltonian.

Finally, in Section V, we apply our general prescription for the study of possible time evolution in the context of $\kappa$-deformed spacetime symmetries. In this case we obtain that the only physically acceptable time evolution is again not described by an equation of the Lindblad type, but by a von Neumann equation governed, this time, by a deformed effective Hamiltonian. The final Section VI contains a summary of the results presented and an outlook for future work.

\section{Time evolution in quantum mechanics}

%\subsection{Standard quantum mechanics}
In ordinary quantum mechanics, the time evolution of the state of a system described by a density matrix $\rho$ is determined by the Hamiltonian $H$ of the system through the von Neumann equation
\begin{equation}
    \partial_t\rho=-i[H,\rho].
    \label{vonn}
\end{equation}
It can be easily shown that the dynamics defined by \eqref{vonn} automatically ensures that positivity, trace and purity of the density matrix are preserved. Trace preservation $\mathrm{Tr}\,\rho(t) = 1$ and positivity $\langle\psi|\rho(t)|\psi\rangle \ge 0$ for any pure state $|\psi\ra$ at any time $t$, are part of the very definition of $\rho$ as a valid representative of a state, since they are statements concerning the probabilistic nature of the information contained in a quantum state. By contrast, the preservation of purity is not a kinematical requirement but a dynamical property, and it holds only for closed systems evolving unitarily.

If we allow the quantum system to interact with an environment, such coupling, under time evolution, will generate entanglement between the system and the external degrees of freedom of the environment. From an observer with exclusive access to the system, an initial pure state will then evolve into a mixed state at later times. The time evolution of an open quantum system interacting with an environment is governed by a generalization of the von Neumann equation \eqref{vonn}, in which purity is no longer preserved. In the general case of Markovian dynamics, such evolution takes the form of a Lindblad equation \cite{Lindblad:1975ef,Gorini:1975nb}
\begin{equation}
    \partial_t\rho=-i[H,\rho]+\sum_{ij}c_{ij}\Big(F_i\rho F_j^\dagger-\frac{1}{2}\{F_j^\dagger F_i,\rho\}\Big),
    \label{Lindb1}
\end{equation}
where $F_i$ are Lindblad (or jump) operators, $c_{ij}$ is a positive semidefinite matrix, known as Kossakowski matrix, and the resulting dynamical map is completely positive and trace preserving. 

The evolution equation \eqref{Lindb1} was re-discovered in \cite{Banks:1983by} in the search for a differential form of Hawking's superscattering operator, the operator describing non-unitary quantum evolution in the presence of black holes \cite{Hawking:1976ra}. In this context, the generalized evolution equation \eqref{Lindb1} plays the role of a {\it fundamental} evolution equation: rather than emerging from the interaction of the system with an environment, it represents an intrinsic feature beyond standard quantum mechanics due to quantum gravitational effects \cite{Hawking:1982dj}. The physical viability of such a generalized evolution in a quantum field theoretic framework was questioned by the authors of \cite{Banks:1983by} and later critically re-examined in \cite{Srednicki:1992bp}. In \cite{Arzano:2014cya} it was suggested by one of the present authors that, in the framework of Hopf algebra deformations of spacetime symmetries associated with spacetime non-commutativity, it could be possible to realize, at a fundamental level, generalized evolution equations of the form \eqref{Lindb1} evading the no-go results of \cite{Banks:1983by}. The main idea at the basis of the proposal in \cite{Arzano:2014cya} is that the Hamiltonian $H$ of a quantum system is, in general, an element of a Lie algebra, and the equation \eqref{vonn} can be understood in terms of the {\it adjoint action} of the generator of time evolution $H$ on a generic operator $\mathcal{O}$ on the Hilbert space of the system 
\begin{equation}\label{adjointactiondef}
    \mathrm{ad}_H(\mathcal{O}) = [H,\mathcal{O}]\,, %\qquad X,Y \in \mathfrak{g}.
\end{equation}
so that the von Neumann equation \eqref{vonn} is simply the adjoint action of $H$ on the density operator
\begin{equation}
    \partial_t \rho = -i\mathrm{ad}_H(\rho)\,.
\end{equation}
Even though at first sight this can appear as a rather trivial observation, it has deep representation-theoretic roots which have far reaching consequences when we start exploring mathematical structures beyond ordinary Lie algebra, such as in the framework of non-trivial Hopf algebras.

The key point is that deformed symmetries lead to a modification of the adjoint action \eqref{adjointactiondef} and, in turn, of the commutator between generic observables. In a standard Lie algebra framework, the action of a generator $X$ on states belonging to the tensor product $\mathcal{H}\otimes\mathcal{H}$ of representations on a Hilbert space $\mathcal{H}$ is governed by the Leibniz rule, encoded by the {\it coproduct}
\begin{equation}
\Delta X = X \otimes \mathbbm{1} + \mathbbm{1} \otimes X,
\label{copro1}
\end{equation}
that ensures additivity of the action of $X$ on multi-systems states. When passing to the dual space $\mathcal{H}^*$, the action of generators on dual bra states must be defined in a way that preserves the natural pairing (inner product) with kets. Given a representation of the Lie algebra on states $|\psi\rangle \in \mathcal{H}$, the action on dual states $\langle\phi| \in \mathcal{H}^*$ is fixed by the requirement
\begin{equation}
( X^{*}\langle\phi|)\,|\psi\rangle
=
-\,\langle\phi|\,( X|\psi\rangle),
\end{equation}
which ensures that the pairing $\langle\phi|\psi\rangle$ is invariant under symmetry transformations. Equivalently, in the language of Hopf algebras, the dual action is implemented via the {\it antipode map} $S$, which acts on generators according to
\begin{equation}
( X^{*}\langle\phi|)\,|\psi\rangle
=
\,\langle\phi|\,( S(X)|\psi\rangle),
\end{equation}
so that for ordinary (undeformed) Lie algebras one has
\begin{equation}
    S(X)= -X\,.
    \label{antip1}
\end{equation}
While the coproduct map \eqref{copro1} encodes additivity of quantum numbers, the antipode reproduces the familiar minus sing required to define the ``inverse element" under addition. Notice how the antipode can be seen as connecting the action of a generator $X$ {\it from the right} in terms of the adjoint $X^{\dagger}$ with the action from the left given by the dual representation
\begin{equation}
    ( X^{*}\langle\phi|)
= \,\langle\phi|S(X))^{\dagger},
\end{equation}
which for hermitian generators of a Lie algebra reduces to
\begin{equation}
    ( X^{*}\langle\phi|)
= - \langle\phi|X\,.
\label{sdag}
\end{equation}
From the point of view of the Hopf algebra maps just discussed, it is easy to see how the adjoint action in terms of a commutator is the natural one to consider when the hermitian generator $X$ is acting on a density matrix (and on any operator which admits a spectral decomposition). To understand this, it suffices to consider the action of $X$ on the projector state $\rho_\psi=|\psi\ra\la \psi|$, the density matrix for the pure state $|\psi\ra$. We have
\be 
X(\rho_\psi) = X(|\psi\ra \la \psi|) =  X(|\psi\ra) \la \psi| + |\psi\ra X^*(\la \psi|) = X |\psi\ra \la \psi| - |\psi\ra \la \psi| X
= [X, \rho_\psi]\,
\ee
where for the second equality we used the coproduct map \eqref{copro1} and in the third the antipode map \eqref{antip1} and the identity \eqref{sdag}. We thus see that the familiar adjoint action in terms of a commutator can be decomposed in terms of the coproduct and antipode as 
\begin{equation}\label{adjconvention}
\mathrm{ad}_X(\rho) := (\mathrm{id} \otimes S)\Delta X \diamond\rho,
\end{equation}
where the operation $\diamond$ is defined by
\begin{equation}
    (A\otimes B)\diamond\rho=A\rho B.
\end{equation}
When dealing with generators $X$ belonging to quantum groups, coproducts and antipodes deviate from their standard forms  $\Delta X = X \otimes \mathbbm{1} + \mathbbm{1}\otimes X$ and $S(X)=-X$ and, correspondingly, the adjoint action \eqref{adjconvention} will no longer reproduce a commutator. The focus of this work will be on exploring the physical viability of these deformed adjoint actions for generalizing the standard von Neumann evolution equation \eqref{vonn} of ordinary quantum mechanics.

Below, we introduce the specific mathematical model of quantum deformations of the Lie algebra $\mathfrak{su}(2)$ which will be the subject of our analysis. 

%\section{Deformation models within the symmetrized prescription}\label{deformedmodels}

\section{$q$-deformed quantum evolution}

\subsection{The $q$-deformed $\mathfrak{su}(2)$ Lie algebra}
\label{subsec:suq2_evolutor}

The $q$-deformed $\mathfrak{su}(2)$ Lie algebra we consider is technically a deformation of its universal enveloping algebra $U(\mathfrak{su}(2))$ denoted by $U_q(\mathfrak{su}(2))$ \cite{biedenharn1995quantum}. In this model the standard Hopf algebra structures of $U(\mathfrak{su}(2))$ in terms of ordinary coproducts and antipodes \eqref{copro1} and \eqref{antip1} are altered, together with the commutators, by terms depending on a {\it deformation parameter} $q\in \mathbb{R}$ which in the limit $q\rightarrow 1$ reproduces the ordinary Hopf algebra structure of $U(\mathfrak{su}(2))$. We adopt here the conventions of \cite{Ballesteros:1998yf}. The algebra is generated by the operators $J_\pm$ and $J_z$
\begin{equation}
[J_z, J_\pm] = \pm J_\pm, \quad [J_+, J_-] = [2J_z]_q,
\end{equation}
with $[x]_q = \frac{q^x - q^{-x}}{q - q^{-1}}$. The non-trivial coproducts and antipodes are given by 
\begin{equation}
\Delta(J_\pm) = J_\pm \otimes q^{-J_z} + q^{J_z} \otimes J_\pm, \quad
\Delta(J_z) = J_z \otimes 1 + 1 \otimes J_z, 
\end{equation}
\begin{equation}
S(J_z) = -J_z, \quad S(J_\pm) = -q^{\mp 1} J_\pm.
\end{equation}
In the limit $q \to 1$ these relations reduce to those of the $\mathfrak{su}(2)$ Lie algebra, supplemented by trivial co-products and antipodes. The action on the basis vectors $\lvert l,m\rangle$ of the $(2l+1)$-dimensional representation of $\mathfrak{su}_q(2)$ is:
\begin{equation}
J_\pm |l,m\rangle = \sqrt{[l\mp m]_q[l\pm m+1]_q}\,|l,m\pm 1\rangle, \quad
2J_z|l,m\rangle = 2m|l,m\rangle.
\end{equation}

Note that, in the case of a qubit, i.e. $l=1/2$, the commutators and the action of the generators reduce to their standard form in the corresponding representation, indeed we have
\begin{equation}
\begin{split}    
    &J_+|\hat{\mathbf{z}},+\rangle=0, \qquad J_-|\hat{\mathbf{z}},+\rangle=\sqrt{\left[\frac{1}{2}+\frac{1}{2}\right]_q\left[\frac{1}{2}-\frac{1}{2}+1\right]_q}|\hat{\mathbf{z}},-\rangle=|\hat{\mathbf{z}},-\rangle,\\
    &J_-|\hat{\mathbf{z}},-\rangle=0,\qquad J_+|\hat{\mathbf{z}},-\rangle=\sqrt{\left[\frac{1}{2}+\frac{1}{2}\right]_q\left[\frac{1}{2}-\frac{1}{2}+1\right]_q}|\hat{\mathbf{z}},+\rangle=|\hat{\mathbf{z}},+\rangle,
\end{split}
\end{equation}
\begin{equation}
    [J_+,J_-]=\frac{q^{2J_z}-q^{-2J_z}}{q-q^{-1}}=\begin{pmatrix} 
        \frac{q-q^{-1}}{q-q^{-1}} &0\\0 &\frac{q^{-1}-q}{q-q^{-1}}
    \end{pmatrix}=2J_z.
\end{equation}
Therefore, in the case of a qubit, the deformation affects only coproducts and antipodes, while the commutators remain undeformed (see also the discussion in \cite{Arzano:2025zsp}).

\subsection{Quantum evolution from adjoint action}\label{Hopfevolutionfromadjointaction}
In order to start our exploration of the possibility of departures from ordinary time evolution in quantum mechanics from non-trivial Hopf algebra, we consider the following $q$-deformed Hamiltonian, proposed originally in \cite{Ballesteros:1998yf} to mimic nonlinear optical interactions
\begin{equation}\label{deformedHSUq2}
H =\epsilon_0\, q^{J_z/2}(J_+ + J_-)q^{J_z/2},
\end{equation}
where $\epsilon_0$ is a constant with dimensions of energy which we introduce for dimensional consistency. The coproduct of this Hamiltonian is given by
\begin{equation}
\Delta(H) =\epsilon_0\, ( H \otimes \mathbbm{1} + q^{2J_z} \otimes H).
\end{equation}
The antipode map must satisfy the defining condition characterizing the antipode in Hopf algebras \cite{Majid_1995} 
\begin{equation}\label{anticond}
m(S\otimes\mathrm{id})\Delta(H)=\eta\, \varepsilon(H),
\end{equation}
where $\eta$ is the unit map and $\varepsilon$ is the counit map\footnote{For an algebra $(A,m,\eta)$, the unit $\eta:k\to A$ encodes the existence of an identity element $\mathbbm{1}\in A$ by embedding scalars as $\eta(\lambda)=\lambda\mathbbm{1}$, such as that multiplication with $\mathbbm{1}$ leaves any element of $A$ unchanged. 
Dually, given a coalgebra $(C,\Delta,\varepsilon)$ over a field $k$, the counit $\varepsilon:C\to k$ is the linear map ensuring that the coproduct $\Delta$ admits a well-defined identity, in the sense that composing $\Delta$ with $\varepsilon$ on either tensor factor reproduces the original element, $(\varepsilon\otimes\mathrm{id})\circ\Delta(c)=c=(\mathrm{id}\otimes\varepsilon)\circ\Delta(c)$. This dual role of unit and counit is a standard ingredient in the definition of bialgebras and Hopf algebras \cite{Tjin:1991me}.}.
Since for the generators of $\mathfrak{su}_q(2)$ $\varepsilon(J_\pm)=\varepsilon(J_z)=0$, we have $\varepsilon(H)=0$. From \eqref{anticond} we obtain
\begin{equation}
S(H) = -\epsilon_0q^{-2J_z}H.
\end{equation}
Note that, for \( q = 1 \), the deformed Hamiltonian \( H \) simply reduces to \( 2\epsilon_0\, J_x \). Furthermore, when the deformed Hamiltonian is evaluated in the qubit representation, it reduces to the undeformed form:
\begin{equation}
\begin{split}   
    H = \epsilon_0 q^{J_z/2}(J_+ + J_-)q^{J_z/2}&=\epsilon_0\begin{pmatrix}
        q^{1/4} &0\\0&q^{-1/4}
    \end{pmatrix}\left[\begin{pmatrix}
        0&1\\0&0
    \end{pmatrix}+\begin{pmatrix}
        0&0\\1&0
    \end{pmatrix}\right]\begin{pmatrix}
        q^{1/4} &0\\0&q^{-1/4}
    \end{pmatrix}\\
    &=\epsilon_0\begin{pmatrix} 0&1\\1&0
    \end{pmatrix}=2\epsilon_0J_x. 
\end{split}
\end{equation}

We now assume that the time evolution of the density matrix $\rho$ is determined by the adjoint action \eqref{adjconvention} of $H$ on the density matrix: 
\begin{equation}
    \mathrm{ad}_H(\rho)=H\rho -q^{2J_z}\rho q^{-2J_z}H\,,
\end{equation}
so that
\be
\partial_t\rho = -i\,\mathrm{ad}_H\rho.
\ee

There is an important caveat we should address at this point concerning the viability of this definition of time evolution. Let us consider a generic Hamiltonian containing a term proportional to the identity $\mathbbm{1}$. This choice is legitimate, since the identity simply shifts the energy spectrum without affecting the physical content. We see that, even in the standard undeformed case, the adjoint action produces an additional contribution proportional to $\rho$ itself, generated precisely by the identity component of $H$. In fact, for a generic
\begin{equation}
    H = \mathbbm{1} + \text{other terms}, 
\end{equation}
given the co-product and antipode of the identity operator \cite{Majid_1995}
\begin{equation}
    \Delta(\mathbbm{1}) = \mathbbm{1}\otimes \mathbbm{1}, \qquad 
    S(\mathbbm{1}) = \mathbbm{1},
\end{equation}
one has
\begin{equation}
    \mathrm{ad}_H(\rho) = (id\otimes S)\Delta(H)\,\diamond\,\rho 
    = \rho + \text{other terms},
\end{equation}
and therefore
\begin{equation}
    \partial_t\rho = -i\,\mathrm{ad}_H(\rho) 
    = -i\rho + \text{other terms}.
\end{equation}
The spurious term $-i\rho$ contributes to the trace of the quantum state as $\partial_t Tr(\rho)=-iTr(\rho)$, which integrates to $Tr(\rho(t))=e^{-it}Tr\rho(0)\neq 1$. In this sense, it violates the normalization condition. In order to reproduce the standard von Neumann evolution from the adjoint action, we therefore have to resort to the following symmetrized definition for the time evolution of the density matrix in terms of the adjoint action of the Hamiltonian $H$
\begin{equation}\label{prescription1}
    \partial_t \rho = \frac{1}{2}\big\{\mathrm{ad}_H(\rho) - [\mathrm{ad}_H(\rho)]^\dagger\big\}.
\end{equation}

Returning to the deformed Hamiltonian \eqref{deformedHSUq2}, setting $q = e^{z/2}$ with $z$ real and small, and expanding to first order, with the prescription \eqref{prescription1} one obtains:
\begin{equation}\label{SUq2deformed_mapped}
\begin{split}
\partial_t \rho &= -\frac{i}{2}\left(H\rho -q^{2J_z}\rho q^{-2J_z}H -\rho H + H q^{-2J_z}\rho q^{2J_z}\right)\\
&\simeq -i[H,\rho] - \frac{iz}{2} \{ H, [\rho, J_z] \} + o(z^2).
\end{split}
\end{equation}
We see that, in the evolution equation defined by the prescription \eqref{prescription1}, the modification of the coproduct and antipode introduce new terms beyond the standard commutator $-i[H,\rho]$. Looking at the leading order contribution in $z$ in \eqref{SUq2deformed_mapped}, we see that this contribution cannot be reabsorbed in a way that the right hand side of  \eqref{SUq2deformed_mapped} is written in terms commutator with some effective Hamiltonian $H_{\mathrm{eff}}$ plus $o(z^2)$ terms 
\be
\partial_t \rho =  -i[H_{\mathrm{eff}},\rho] + o(z^2)\,.
\ee
As a consequence, the evolution \eqref{SUq2deformed_mapped}  does not guarantee positivity for all states under time evolution and, in fact, we can explicitly see that the evolution  \eqref{SUq2deformed_mapped} fails to preserve positivity. Let $\rho_0 = \ket{\phi}\bra{\phi}$ be a pure state, and let $\ket{\psi}$ be any vector orthogonal to $\ket{\phi}$, 
\begin{equation} 
\braket{\psi}{\phi} = 0. 
\end{equation} 
For small times $t$, $\rho(t)$ can be expanded as 
\begin{equation}
\rho(t) = \rho_0 + t\,\mathcal{L}(\rho_0) + o(t), \end{equation} 
with $\mathcal{L}(\rho_0)$ given by
\begin{equation} \label{Liouvilliangenerator}
\mathcal{L}(\rho_0) = -i[H,\rho_0] - \frac{i z}{2}\,\{H,[\rho_0,J_z]\}\,. 
\end{equation} 
Taking the expectation value on $\ket{\psi}$, we obtain 
\begin{equation}\label{eq:psi_expansion} \bra{\psi}\rho(t)\ket{\psi} = \bra{\psi}\rho_0\ket{\psi} + t\,\bra{\psi}\mathcal{L}(\rho_0)\ket{\psi} + o(t) = t\,\bra{\psi}\mathcal{L}(\rho_0)\ket{\psi} + o(t), 
\end{equation} 
since $\bra{\psi}\rho_0\ket{\psi} = |\braket{\psi}{\phi}|^2 = 0$. We have
\begin{equation} 
\bra{\psi}\mathcal{L}(\rho_0)\ket{\psi} = - \frac{i z}{2}\,\bra{\psi}\{H,[\rho_0,J_z]\}\ket{\psi}, 
\end{equation}
which depends on matrix elements of the form
$\langle\psi|H|\phi\rangle\langle\phi|J_z|\psi\rangle$ and $\langle\psi|J_z|\phi\rangle\langle\phi|H|\psi\rangle$, which can be made non-zero and with arbitrary sign by a suitable choice of $\ket{\phi}$, $\ket{\psi}$, $H$ and $J_z$ on a finite-dimensional Hilbert space. In particular, one may choose them so that
\begin{equation}
\bra{\psi}\mathcal{L}(\rho_0)\ket{\psi} < 0,
\end{equation}
and therefore even for small $t$ one can have
\begin{equation}
\bra{\psi}\rho(t)\ket{\psi} = t\,\bra{\psi}\mathcal{L}(\rho_0)\ket{\psi} + o(t^2) < 0.
\end{equation}
Let us consider for example the basis states $|\hat{\mathbf{z}},\pm\rangle$, i.e the eigenstates of $J_z$. Let us choose the orthogonal states
\begin{equation}
\ket{\phi} = \ket{\hat{\mathbf{y}},+} =
\frac{\,|\hat{\mathbf{z}},-\rangle - i\,|\hat{\mathbf{z}},+\rangle\,}{\sqrt{2}},
\quad
\ket{\psi} = \ket{\hat{\mathbf{y}},-} =
\frac{\,|\hat{\mathbf{z}},-\rangle + i\,|\hat{\mathbf{z}},+\rangle\,}{\sqrt{2}}.
\end{equation}
so that $\rho_0=\ket{\phi}\bra{\phi}$ and $\braket{\psi}{\phi}=0$. A direct calculation of \eqref{Liouvilliangenerator}, using the qubit representation of the generators, gives

\begin{equation}
\bra{\psi}\mathcal{L}(\rho_0)\ket{\psi} = -\frac{z}{2},
\end{equation}
which is strictly negative for any $z>0$, showing explicitly that the evolution fails to preserve positivity of the density matrix.\\

Finally, let us notice that, in order to have a deformed evolution of the Lindblad type \eqref{Lindb1}, one should consider a real term at leading order in $z$ in \eqref{SUq2deformed_mapped}. Thus, one should require $z\in i\mathbb{R}$. However, this hypothesis spoils the hermiticity of the coproducts of the generators, making the evolution model no longer physically admissible since the quantum numbers associated to generator would add up to complex values for multiparticle states. We are thus led to the conclusion that the deformed Hamiltonian \eqref{deformedHSUq2} cannot define a consistent time evolution for a density matrix under the prescription \eqref{prescription1}. This of course does not mean that it is impossible to define a sensible time evolution using the Hamiltonian \eqref{deformedHSUq2}. In order to push our quest further, in the next subsection we explore the possibility of defining a quantum evolution in terms of the adjoint action of $H$ using more general prescriptions than \eqref{prescription1}.

\subsection{Combining deformed adjoint actions}

From an algebraic perspective, the prescription \eqref{prescription1} for defining an evolution equation via the adjoint action of the Hamiltonian {\it is not} the only one that reduces, in the case of a trivial Hopf algebra, to the standard von Neumann evolution equation. We have, in fact, that in the formal theory of Hopf algebras one can define {\it left} and  {\it right} adjoint actions \cite{Majid_1995} as follows
\begin{equation}\label{LRadj}
\begin{aligned}
&\mathrm{ad^L}_H(\rho)=(id\otimes S)\Delta H\diamond \rho,\\
&\mathrm{ad^R}_H(\rho)=(S\otimes id)\Delta H\diamond \rho.
\end{aligned}
\end{equation}
In the undeformed case, these actions reproduce respectively  $[H,\rho]$ and $-[H,\rho] = [\rho, H]$. We thus see that the most general definition of time evolution can obtained by a linear combination of \eqref{LRadj} and their Hermitian adjoints
\begin{equation}\label{genericcombination}
i\partial_t\rho=\alpha\mathrm{ad^L}_H(\rho)+\beta[\mathrm{ad^L}_H(\rho)]^\dagger+\gamma \mathrm{ad^R}_H(\rho)+\delta [\mathrm{ad^R}_H(\rho) ]^\dagger,
\end{equation}
with real coefficients $(\alpha,\beta,\gamma,\delta)$ satisfying the conditions
\begin{equation}\label{constraint0}
    \alpha+\beta+\gamma+\delta=0, \quad\alpha-\beta-\gamma+\delta=1
\end{equation} 
in order to ensure that in the undeformed case $i\partial_t\rho=[H,\rho]$. The first condition is required to cancel the contribution that would arise from a possible identity term in the Hamiltonian, as discussed in subsection \ref{Hopfevolutionfromadjointaction}. The second condition ensures that the undeformed evolution reduces to the standard von Neumann commutator. In the undeformed case, these two constraints are sufficient to guarantee the physical consistency of the evolution equation, without fixing a unique numerical value for the coefficients.\\ 

Let us now look at the generic combination of adjoint actions \eqref{genericcombination} in order to find out if there is a particular choice that guarantees a viable time evolution for the Hamiltonian \eqref{deformedHSUq2}. Let us first compute the left and right adjoint actions at leading order in $z$:

\begin{equation}\label{firstadjoint}
\begin{aligned}
\mathrm{ad}_H^L(\rho)
&= H\rho + q^{2J_z}\rho S(H) \\
&=\epsilon_0\{ q^{J_z/2}(J_+ + J_-)q^{J_z/2}\rho
   + q^{2J_z}\rho S (q^{J_z/2}(J_+ + J_-)q^{J_z/2})\} \\
&=\epsilon_0\{ e^{zJ_z}(J_+ + J_-)e^{zJ_z}\rho
   + e^{zJ_z}\rho S(e^{zJ_z}(J_+ + J_-)e^{zJ_z})\} \\
&\simeq \epsilon_0\left[(J_+ + J_-),\rho\right] + \epsilon_0 z\Big(
   J_z(J_+ + J_-)\rho
   + (J_+ + J_-)J_z\rho
   - J_z\rho(J_+ + J_-) \\
&  + \rho J_z(J_+ + J_-)
   + \rho(J_+ + J_-)J_z
   + \tfrac{1}{2}\rho(J_+ - J_-)
   \Big)+o(z^2),
\end{aligned}
\end{equation}

\begin{equation}\label{secondadjoint}
\begin{aligned}
    \mathrm{ad}_H^R(\rho)
    &= S(H)\rho + S(q^{2J_z})\rho H \\
    &= \epsilon_0 S(q^{J_z/2}(J_+ + J_-)q^{J_z/2})\rho
       + S(q^{2J_z})\rho H \\
    &=\epsilon_0 S(e^{zJ_z}(J_+ + J_-)e^{zJ_z})\rho
       + S(e^{zJ_z})\rho H \\
    &\simeq -\epsilon_0\left[(J_+ + J_-),\rho\right]
       + \epsilon_0 z\Bigl(
           J_z(J_+ + J_-)\rho
           + (J_+ + J_-)J_z\rho \\
    &      + \rho J_z(J_+ + J_-)
           + \rho(J_+ + J_-)J_z
           - J_z\rho(J_+ + J_-)
           + \tfrac{1}{2}(J_+ - J_-)\rho
         \Bigr)+o(z^2).
\end{aligned}
\end{equation}
Plugging the expressions above in \eqref{genericcombination} we obtain

\begin{equation}\label{11}
\begin{aligned}
i\partial_t\rho 
&=\epsilon_0 (\alpha-\beta-\gamma+\delta)\,\big[(J_+ + J_-),\rho\big] \\
&\quad + \epsilon_0 z\Big\{ 
   (\alpha+\beta+\gamma+\delta)\,\big(J_z(J_+ + J_-)\rho + (J_+ + J_-)J_z\rho\big) \\
&\qquad + (-\alpha-\gamma)\,\big(J_z\rho(J_+ + J_-)\big) \\
&\qquad + (\alpha+\delta+\beta+\gamma)\,\big(\rho J_z(J_+ + J_-)\big) \\
&\qquad + (\alpha+\beta+\gamma+\delta)\,\big(\rho(J_+ + J_-)J_z\big) \\
&\qquad + \tfrac{1}{2}(\alpha-\delta)\,\rho(J_+ - J_-) 
          + \tfrac{1}{2}(\gamma-\beta)\,(J_+ - J_-)\rho
   \Big\}+o(z^2).
\end{aligned}
\end{equation}

A closer inspection of this equation shows that the contribution of order z contains terms such as $J_z(J_+ + J_-)\rho$, $(J_+ + J_-)J_z\rho$, $\rho J_z(J_+ + J_-)$ and $\rho(J_+ + J_-)J_z$, all appearing with the same sign and multiplied by the same linear combination of coefficients. Such terms cannot be recast into a pure commutator. Moreover, the product term $J_z\rho(J_+ + J_-)$ has no partner with which it could combine into a commutator. Requiring these terms to vanish, together with the condition that the undeformed term appears with unit prefactor, leads to the system
\begin{equation}
\begin{cases}
\alpha-\beta-\gamma+\delta=1,\\
\alpha+\beta+\gamma+\delta = 0,\\
\alpha+\gamma=0,\\
\beta+\delta= 0,
\end{cases}
\end{equation}
whose unique solution is
\begin{equation}
\label{1q1q1q1q}
\alpha = \frac14,\qquad
\beta = -\frac14,\qquad
\gamma = -\frac14,\qquad
\delta = \frac14.
\end{equation}
This is the only choice of coefficients that guarantees a von Neumann-type evolution, thereby making the model physically consistent. We conclude that the prescription
\begin{equation}\label{prescriptiongeneral}
\begin{aligned}
    i\partial_t\rho&=\frac{1}{4}\left\{\mathrm{ad}_H^L(\rho)-[\mathrm{ad}_H^L(\rho)]^\dagger-\mathrm{ad}^R_H(\rho)+[\mathrm{ad}^R_H(\rho)]^\dagger\right\}%=[(J_++J_-),\rho].
\end{aligned}
\end{equation}
is the only one that leads to a viable evolution equation of the von Neumann form 
\begin{equation}
    \partial_t\rho=-i[\epsilon_0(J_++J_-),\rho]=-i[H,\rho],
\end{equation}
governed by the {\it undeformed} Hamiltonian 
\begin{equation}
    H=\epsilon_0(J_++J_-).
\end{equation}
The above analysis has been carried out explicitly for the deformed Hamiltonian \eqref{deformedHSUq2}, which in qubit representation is proportional to $J_x$, but the result extends to any Hamiltonian built from the Pauli basis. First, replacing $J_+ + J_-$ with $J_+ - J_-$ in the derivation shows that the deformation terms generated by $H \propto J_y$ have exactly the same operator structure and depend on $(\alpha,\beta,\gamma,\delta)$ through the same linear combinations as in the
$J_x$ case. The only change is the replacement of $H$ in \eqref{firstadjoint}–\eqref{secondadjoint}. Hence, the constraints required to eliminate non–commutator terms are the same for $J_x$ and $J_y$. Secondly, the coproduct of $J_z$ in this model is undeformed. As a consequence, for $H\propto J_z$ the left and right adjoint actions reduce exactly to $\mathrm{ad}^L_H(\rho)=[H,\rho]$ and $\mathrm{ad}^R_H(\rho)=-[H,\rho]$, with no deformation terms. Taken together, these observations show that the condition \eqref{1q1q1q1q} obtained in the $J_x$ case are universal for any qubit Hamiltonian in the deformed $\mathfrak{su}(2)$ framework considered. 

We conclude this section by analyzing the case of {\it complex} linear combinations of left an right adjoint action and their Hermitian adjoints. Indeed, choosing $\alpha,\beta,\gamma,\delta \in \mathbb{C}$, could allow for an additional real contribution to the evolution equation without spoiling the hermiticity of the Hamiltonian and thus lead to the possibility of a Lindblad-like evolution equation. Let us set

\begin{equation}\label{decomposition}
\begin{split}
    &\alpha=\alpha_0+i\alpha_1,\\
    &\beta=\beta_0+i\beta_1,\\
    &\gamma=\gamma_0+i\gamma_1,\\
    &\delta=\delta_0+i\delta_1.
\end{split}
\end{equation}
For the evolution equation under the general prescription \eqref{genericcombination} we have

\begin{equation}
\begin{aligned}
i\partial_t\rho 
&= \epsilon_0(\alpha_0-\beta_0-\gamma_0+\delta_0)\,[J_+ + J_-,\rho]
   + i\epsilon_0(\alpha_1-\beta_1-\gamma_1+\delta_1)\,[J_+ + J_-,\rho]\\
&\quad + \epsilon_0 z\Big\{ 
   (\alpha_0+\beta_0+\gamma_0+\delta_0)\,\big(J_z(J_+ + J_-)\rho + (J_+ + J_-)J_z\rho\big) \\
&\qquad + i(\alpha_1+\beta_1+\gamma_1+\delta_1)\,\big(J_z(J_+ + J_-)\rho + (J_+ + J_-)J_z\rho\big) \\
&\qquad + (-\alpha_0-\gamma_0)\,J_z\rho(J_+ + J_-)
          + i(-\alpha_1-\gamma_1)\,J_z\rho(J_+ + J_-) \\
&\qquad + (\alpha_0+\beta_0+\gamma_0+\delta_0)\,\rho J_z(J_+ + J_-)
          + i(\alpha_1+\beta_1+\gamma_1+\delta_1)\,\rho J_z(J_+ + J_-) \\
&\qquad + (\alpha_0+\beta_0+\gamma_0+\delta_0)\,\rho(J_+ + J_-)J_z
          + i(\alpha_1+\beta_1+\gamma_1+\delta_1)\,\rho(J_+ + J_-)J_z \\
&\qquad + \tfrac{1}{2}(\alpha_0-\delta_0)\,\rho(J_+ - J_-)
          + \tfrac{1}{2}(\gamma_0-\beta_0)\,(J_+ - J_-)\rho \\
&\qquad + \tfrac{i}{2}(\alpha_1-\delta_1)\,\rho(J_+ - J_-)
          + \tfrac{i}{2}(\gamma_1-\beta_1)\,(J_+ - J_-)\rho
   \Big\}+o(z^2).
\end{aligned}
\end{equation}
For $\alpha_0,\beta_0,\gamma_0,\delta_0$ we must again require $\alpha_0=-\beta_0=-\gamma_0=\delta_0=\tfrac{1}{4}$ to allow for a correct von Neumann undeformed limit. For the imaginary part, it is instead required that 
$\alpha_1-\beta_1-\gamma_1+\delta_1=0$ in order to preserve the undeformed evolution term. The equation reduces to
\begin{equation}
\begin{aligned}
i\partial_t\rho 
&= [H,\rho] 
   + i \epsilon_0 z \Big\{ 
      (-\alpha_1-\gamma_1)\,J_z\rho(J_+ + J_-)\\
&\qquad + 2(\alpha_1+\delta_1)\,\Big(
          J_z(J_+ + J_-)\rho
          + \rho J_z(J_+ + J_-)
          + \rho(J_+ + J_-)J_z
      \Big) \\
&\qquad + \tfrac{1}{2}(\alpha_1-\delta_1)\,\rho(J_+ - J_-)
          + \tfrac{1}{2}(\gamma_1-\beta_1)\,(J_+ - J_-)\rho
   \Big\}+o(z^2).
\end{aligned}
\end{equation}
It is evident that this equation cannot be reduced to an effective Lindblad dynamics, since it cannot be recast into the Lindblad form \eqref{Lindb1}. Indeed, for each term of the type $F_i \rho F_j^\dagger$, there must appear the corresponding anticommutator contributions $F_j^\dagger F_i \rho$ and $\rho F_j^\dagger F_i$ with fixed relative coefficients. In our case, the terms $J_z\rho(J_+ + J_-)$, $\rho J_z(J_+ + J_-)$ and $\rho(J_+ + J_-)J_z$, do not recombine in the right way because of their signs and prefactors. More importantly, there are unpaired combinations, such as $\rho(J_++J_-)J_z$, $\rho(J_+-J_-)$, $(J_+-J_-)\rho$, appearing without the associated terms that a Lindblad-type equation would exhibit. The simultaneous mismatch of relative coefficients for the potentially pairable terms and the presence of such unpaired contributions show that the evolution cannot be recast into Lindblad form. 

It is interesting to mention that, considering an imaginary deformation parameter mapping $z\to iz$, under specific approximations the evolution equation reproduces a Redfield-like form \cite{5392713}:
\begin{equation}
\dot{\rho}(t)= -i[H_S,\rho(t)]- \sum_{\alpha,\beta} \int_0^{\infty} d\tau \Big(C_{\alpha\beta}(\tau)[A_\alpha, A_\beta(-\tau)\rho(t)]+ C_{\beta\alpha}(-\tau)[A_\alpha, \rho(t) A_\beta(-\tau)]\Big),
\end{equation} 
where
\begin{equation}
A_\beta(-\tau)=e^{-iH_S\tau}A_\beta e^{iH_S\tau}, \qquad
C_{\alpha\beta}(\tau)=\mathrm{Tr}_E[B_\alpha(\tau)B_\beta\rho_E].
\end{equation}
In particular, under the approximations
\begin{enumerate}
\item Instantaneous correlations: 
\begin{equation}
C_{\alpha\beta}(\tau)\approx \gamma\,\delta(\tau), \qquad
C_{\beta\alpha}(-\tau)\approx \gamma\,\delta(\tau),
\end{equation}

\item Short times: 
\begin{equation}
A_\beta(-\tau)\approx A_\beta - i\tau[H_S,A_\beta],
\end{equation}
\end{enumerate}
and Formally identifying $A_\alpha \rightarrow H$, $A_\beta \rightarrow J_z$, $\gamma \rightarrow z/2$, one finds:
\begin{equation}
\dot{\rho}(t)= -i[H_S,\rho(t)]-\frac{z}{2}\left(
[\rho, J_z]\,H + H\,[J_z,\rho]
\right),
\end{equation}
which is the equation emerging in the imaginary deformation case. It is important to stress that the Redfield equation fails to guarantee complete positivity of the state evolution at all times. Moreover, the map $z\to iz$ spoils the hermiticity of the coproducts of the generators, making the model unphysical. Nevertheless, this feature can be relevant for possible effective models describing information scrambling, as discussed in \cite{Tripathy:2025uby}, particularly in scenarios involving deformed symmetry structures, as suggested in \cite{Arzano:2025zsp}.

\section{Time evolution under a general deformation of $\mathfrak{su}(2)$}\label{dimostrazione}
In the previous section we showed that for the quantum group $U_q(\mathfrak{su}(2))$, the most widely studied Hopf algebra deformation of $\mathfrak{su}(2)$, the only prescription which leads to a physically acceptable time evolution equation in terms of deformed adjoint actions is \eqref{prescriptiongeneral}. Significantly, the evolution equation obtained through this prescription is given by the standard von Neumann equation. 

In this section we introduce the most general Hopf-algebra deformations of $\mathfrak{su}(2)$ with undeformed algebraic sector and deformed coproducts {\it at leading order in the deformation parameter}. We will see that, also in this case,  the only prescription leading to a physically viable evolution equation is \eqref{prescriptiongeneral} which, again, leads to a von Neumann–type time evolution.

\subsection{The model}

Let us consider the following leading order deformation of the coproducts of the generators of $\mathfrak{su}(2)$

\begin{equation}
\Delta(J_k)
= J_k \otimes \mathbbm{1} + \mathbbm{1} \otimes J_k
\;+\;
h \sum_{i,j \in \{+,-,z\}} c^{(k)}_{ij}\, J_i \otimes J_j;
\quad  k \in \{+,-,z\}\,
\end{equation}
where $h$ is a dimensionless deformation parameter.\\
As stressed above, we restrict to the case where the commutators of the generators are undeformed
\begin{equation}
\begin{split}
&[J_+,J_-]=2J_z,\quad [J_z,J_+]=J_+,\quad [J_z,J_-]=-J_-.
\end{split}
\end{equation}
This condition allows us to use the Pauli representation for the single particle generators. Imposing the coproduct homomorphism condition:
\begin{equation}
\begin{split}
&\Delta([J_+,J_-]) = [\Delta(J_+),\Delta(J_-)],\\
&\Delta([J_z,J_+]) = [\Delta(J_z),\Delta(J_+)],\\
&\Delta([J_z,J_-]) = [\Delta(J_z),\Delta(J_-)]\,,
\end{split}
\end{equation}
we obtain a linear system of $19$ independent equations involving $27$ unknown coefficients $c^{(k)}_{ij}$. Specifically, for the coproduct of \(J_z\) we have
\begin{equation}\label{copz1}
\begin{cases}
c^{(z)}_{+-} = 0,\\
c^{(z)}_{-+} = 0,\\
c^{(z)}_{zz} = 0,
\end{cases}
\end{equation}
for the coproduct of \(J_+\)
\begin{equation}\label{cop+1}
\begin{cases}
c^{(+)}_{++} = -\big(c^{(z)}_{+z} + c^{(z)}_{z+}\big),\\
c^{(+)}_{+-} = \; c^{(z)}_{z-},\\
c^{(+)}_{-+} = \; c^{(z)}_{-z},\\
c^{(+)}_{--} = 0,\\
c^{(+)}_{-z} = -\dfrac{1}{4}\,c^{(z)}_{--},\\
c^{(+)}_{z-} = -\dfrac{1}{4}\,c^{(z)}_{--},\\
c^{(+)}_{zz} = -\dfrac{1}{2}\,\big(c^{(z)}_{-z} + c^{(z)}_{z-}\big).
\end{cases}
\end{equation}
and, finally, for the coproduct of \(J_-\)
\begin{equation}\label{cop-1}
\begin{cases}
c^{(-)}_{++} = 0,\\
c^{(-)}_{+-} = \; c^{(z)}_{+z},\\
c^{(-)}_{+z} = -\dfrac{1}{4}\,c^{(z)}_{++},\\
c^{(-)}_{-+} = \; c^{(z)}_{z+},\\
c^{(-)}_{--} = -\big(c^{(z)}_{-z} + c^{(z)}_{z-}\big),\\
c^{(-)}_{-z} = -\,c^{(+)}_{z+},\\
c^{(-)}_{z+} = -\dfrac{1}{4}\,c^{(z)}_{++},\\
c^{(-)}_{z-} = -\,c^{(+)}_{+z},\\
c^{(-)}_{zz} = -\dfrac{1}{2}\,\big(c^{(z)}_{+z} + c^{(z)}_{z+}\big).
\end{cases}
\end{equation}

The general solution depends on $8$ free parameters that, for definiteness, we choose to be: $c^{(z)}_{++},\; c^{(z)}_{+z},\; c^{(z)}_{--},\; c^{(z)}_{-z},\; c^{(z)}_{z+},\; c^{(z)}_{z-},\; c^{(+)}_{+z},\; c^{(+)}_{z+}$. Imposing the Hermiticity condition $\Delta J_z = (\Delta J_z)^\dagger$, $\Delta J_+ = (\Delta J_-)^\dagger$ and $\Delta J_- = (\Delta J_+)^\dagger$ yields the following extra constraints
\begin{equation}
c^{(z)}_{++} = {c^{(z)}_{--}}^*, \qquad
c^{(z)}_{+z} = {c^{(z)}_{-z}}^*, \qquad
c^{(z)}_{z+} = {c^{(z)}_{z-}}^*,\qquad c^{(z)}_{zz}={c^{(z)}_{zz}}^*.
\end{equation}
\begin{equation} 
\begin{aligned} 
c^{(-)}_{++} &= {c^{(+)}_{--}}^*, &\qquad c^{(-)}_{--} &= {c^{(+)}_{++}}^*, &\qquad c^{(-)}_{+-} &= {c^{(+)}_{-+}}^*, &\qquad c^{(-)}_{-+} &= {c^{(+)}_{+-}}^*, \\ c^{(-)}_{+z} &= {c^{(+)}_{-z}}^*, &\qquad c^{(-)}_{-z} &= {c^{(+)}_{+z}}^*, &\qquad c^{(-)}_{z+} &= {c^{(+)}_{z-}}^*, &\qquad c^{(-)}_{z-} &= {c^{(+)}_{z+}}^*, \\  c^{(-)}_{zz} &= {c^{(+)}_{zz}}^*. 
\end{aligned} 
\end{equation}
Using these constraints we obtain the following expressions for the coproducts:
\begin{equation}
\begin{split}
\Delta(J_z) &= J_z \otimes \mathbbm{1} + \mathbbm{1} \otimes J_z \\
&+ h\Big[
c^{(z)}_{++}\,J_+ \otimes J_+ 
+ c^{(z)}_{+z}\,J_+ \otimes J_z
+ c^{(z)}_{--}\,J_- \otimes J_- \\
& + c^{(z)}_{-z}\,J_- \otimes J_z
+ c^{(z)}_{z+}\,J_z \otimes J_+
+ c^{(z)}_{z-}\,J_z \otimes J_-
\Big],
\end{split}
\end{equation}
\begin{equation}
\begin{split}
\Delta(J_+) &= J_+ \otimes \mathbbm{1} + \mathbbm{1} \otimes J_+ \\
& + h\Big[
c^{(+)}_{+z}\,J_+ \otimes J_z
+ c^{(+)}_{z+}\,J_z \otimes J_+ \\
&-\big(c^{(z)}_{+z} + c^{(z)}_{z+}\big)\,J_+ \otimes J_+ 
+ c^{(z)}_{z-}\,J_+ \otimes J_- 
+ c^{(z)}_{-z}\,J_- \otimes J_+ \\
&-\tfrac14\,c^{(z)}_{--}\,J_- \otimes J_z
- \tfrac14\,c^{(z)}_{--}\,J_z \otimes J_- -\tfrac12\big(c^{(z)}_{-z} + c^{(z)}_{z-}\big)\,J_z \otimes J_z
\Big],
\end{split}
\end{equation}
\begin{equation}
\begin{split}
\Delta(J_-) &= J_- \otimes \mathbbm{1} + \mathbbm{1} \otimes J_- \\
&+ h\Big[
-\tfrac14\,c^{(z)}_{++}\,J_+ \otimes J_z
-\tfrac14\,c^{(z)}_{++}\,J_z \otimes J_+ \\
& + c^{(z)}_{+z}\,J_+ \otimes J_- 
+ c^{(z)}_{z+}\,J_- \otimes J_+ -\big(c^{(z)}_{-z} + c^{(z)}_{z-}\big)\,J_- \otimes J_- \\
& -\,c^{(+)}_{z+}\,J_- \otimes J_z
-\,c^{(+)}_{+z}\,J_z \otimes J_- -\tfrac12\big(c^{(z)}_{+z} + c^{(z)}_{z+}\big)\,J_z \otimes J_z
\Big].
\end{split}
\end{equation}
Let us now derive the leading order deformations for the  antipodes $S(J_k)$ starting from the coproducts $\Delta(J_k)$ we just determined. We postulate the linear ansatz:
\begin{equation}
S(J_k) \;=\; -J_k \;+\; h\big( s^{(k)}_+ J_+ + s^{(k)}_- J_- + s^{(k)}_z J_z \big),
\end{equation}
where the coefficients $s^{(k)}_i$ are to be determined as functions of the $c^{(k)}_{ij}$. The Hopf algebra consistency condition is given by
\begin{equation}\label{antipode-conds}
m(S\otimes \mathrm{id})\,\Delta(J_k)=0,
\quad \forall\,k\in\{+,-,z\}\,,
\end{equation}
which explicitly reads
\begin{equation}
\begin{split}
m(S\otimes \mathrm{id})\,\Delta(J_k)
&= S(J_k)\cdot \mathbbm{1} + \mathbbm{1}\cdot J_k
+ h\sum_{i,j} c^{(k)}_{ij}\, S(J_i)\,J_j\\
&= \big(-J_k + h\sum_i s^{(k)}_i J_i\big) + J_k
+ h\sum_{i,j} c^{(k)}_{ij}\,(-J_i)J_j + O(h^2)\\
&= h\left[\sum_i s^{(k)}_i J_i - \sum_{i,j} c^{(k)}_{ij}\,J_iJ_j\right] + o(h^2)\,,
\end{split}
\end{equation}
leading to the condition 
\begin{equation}\label{conditionn}
\sum_i s^{(k)}_i J_i \;=\; \sum_{i,j} c^{(k)}_{ij}\,J_iJ_j.
\end{equation}
Solving this equation we obtain the following relations for the coefficients $s^{(k)}_i$:

\begin{equation}
\begin{cases}
s^{(z)}_+ = -c^{(z)}_{+z} + c^{(z)}_{z+},\\
s^{(z)}_- = c^{(z)}_{-z} - c^{(z)}_{z-},\\
s^{(z)}_z = 0
\end{cases}
\end{equation}

\begin{equation}
\begin{cases}
s^{(+)}_+ = -\,c^{(+)}_{+z} + c^{(+)}_{z+},\\
s^{(+)}_- = 0,\\
s^{(+)}_z = \tfrac{1}{2}\big(c^{(z)}_{-z} - c^{(z)}_{z-}\big)
\end{cases}
\end{equation}

\begin{equation}
\begin{cases}
s^{(-)}_+ = 0,\\
s^{(-)}_- = c^{(+)}_{z+} - c^{(+)}_{+z},\\
s^{(-)}_z = \tfrac{1}{2}\big(c^{(z)}_{z+} - c^{(z)}_{+z}\big).
\end{cases}
\end{equation}

The resulting antipodes are given by

\begin{equation}\label{antipodesigmazconstrained}
S(J_z) = -J_z
+ h\Big[\,(-c^{(z)}_{+z} + c^{(z)}_{z+})\,J_+
+(c^{(z)}_{-z} - c^{(z)}_{z-})\,J_- \Big],
\end{equation} 

\begin{equation}
S(J_+) = -J_+
+ h\Big[\,(-c^{(+)}_{+z}+c^{(+)}_{z+})\,J_+
+\tfrac{1}{2}(c^{(z)}_{-z}-c^{(z)}_{z-})\,J_z \Big],
\end{equation}

\begin{equation}
    S(J_-) = -J_- + h\Big[\,(c^{(+)}_{z+}-c^{(+)}_{+z})\,J_-
+\tfrac{1}{2}(c^{(z)}_{z+}-c^{(z)}_{+z})\,J_z \Big].
\end{equation}

\subsection{General adjoint action}\label{general adjoint action}\label{VB}

We will now study the time evolution equations obtained under the prescription \eqref{genericcombination} to check which choices of $\alpha,\beta,\gamma,\delta$ lead to a physically admissible evolution. For definiteness let us consider the following Hamiltonian:
\begin{equation}\label{hamiltonianmodel}
H=\epsilon_0J_z.
\end{equation}
%The coproduct and the antipode of $H$ are
%\begin{equation}
 % \Delta H=\epsilon_0 \left(J_z\otimes \mathbbm{1}+\mathbbm{1}\otimes J_z\right),
%\end{equation}
%\begin{equation}
%    S(H)=-\epsilon_0 J_z.
%\end{equation}
From the results of the previous section we have that the coproduct and antipode of the Hamiltonian are given by
\begin{equation}
\begin{split}
    \Delta H = \epsilon_0 \Big[ 
    & J_z \otimes \mathbbm{1} + \mathbbm{1} \otimes J_z \\
    & + h \Big(
        c^{(z)}_{++}\, J_+ \otimes J_+ 
        + c^{(z)}_{+z}\, J_+ \otimes J_z
        + c^{(z)}_{--}\, J_- \otimes J_- \\
    & \quad + c^{(z)}_{-z}\, J_- \otimes J_z
        + c^{(z)}_{z+}\, J_z \otimes J_+
        + c^{(z)}_{z-}\, J_z \otimes J_-
    \Big)
\Big],
\end{split}
\end{equation}
and
\begin{equation}
    S(H)=\epsilon_0\left\{-J_z+h[(-c_{+z}^{(z)}+c_{z+}^{(z)})J_++(c_{-z}^{(z)}-c_{z-}^{(z)})J_-]\right\}.
\end{equation}
From these expressions we can compute the left adjoint action 
\begin{equation}
\begin{split}
    ad_H^L(\rho) 
    &= (id \otimes S)\,\Delta H \diamond \rho \\
    &\simeq \epsilon_0 \Big[
        J_z \rho - \rho J_z + h\Big(
        (-c^{(z)}_{+z} + c^{(z)}_{z+})\, \rho J_+
        + (c^{(z)}_{-z} - c^{(z)}_{z-})\, \rho J_- 
    \Big) \\
    &\quad + h\Big(
        - c^{(z)}_{++}\, J_+ \rho J_+ 
        - c^{(z)}_{+z}\, J_+ \rho J_z
        - c^{(z)}_{--}\, J_- \rho J_- \\
    &\qquad\quad
        - c^{(z)}_{-z}\, J_- \rho J_z
        - c^{(z)}_{z+}\, J_z \rho J_+
        - c^{(z)}_{z-}\, J_z \rho J_-
    \Big)
    \Big] + o(h^2),
\end{split}
\end{equation}
while for the right adjoint action we have
\begin{equation}
\begin{split}
    ad_H^R(\rho) 
    &= (S \otimes id)\,\Delta H \diamond \rho \\
    &= \epsilon_0 \Big[
        \rho J_z - J_z \rho+ h \Big(
        (-c^{(z)}_{+z} + c^{(z)}_{z+})\, J_+ \rho
        + (c^{(z)}_{-z} - c^{(z)}_{z-})\, J_- \rho
    \Big) \\
    &\quad + h \Big(
        - c^{(z)}_{++}\, J_+ \rho J_+ 
        - c^{(z)}_{+z}\, J_+ \rho J_z
        - c^{(z)}_{--}\, J_- \rho J_- \\
    &\qquad\quad
        - c^{(z)}_{-z}\, J_- \rho J_z
        - c^{(z)}_{z+}\, J_z \rho J_+
        - c^{(z)}_{z-}\, J_z \rho J_-
    \Big)
    \Big] + o(h^2).
\end{split}
\end{equation}
The most general evolution equation  obtained from the prescription \eqref{genericcombination} is then given by
\begin{equation}
\begin{split}
    i\partial_t\rho
    &=\epsilon_0\Big\{(\alpha-\beta-\gamma+\delta)(J_z\rho-\rho J_z)\\
    &+h\Big[(-c_{+z}^{(z)}+c_{z+}^{(z)})((\alpha-\delta)\rho J_++(\gamma-\beta)J_+\rho) + (-c_{-z}^{(z)}+c_{z-}^{(z)})((\delta-\alpha)\rho J_- +(\beta-\gamma)J_-\rho)\\
    &- c_{++}^{(z)}\, (\alpha + \beta + \gamma + \delta)\, J_+ \rho J_+ - c_{--}^{(z)}\, (\alpha + \beta + \gamma + \delta)\, J_- \rho J_- 
      - \big( c_{+z}^{(z)}(\alpha + \gamma) + c_{z+}^{(z)}(\beta + \delta) \big)\, J_+ \rho J_z\\
      &- \big( c_{-z}^{(z)}(\alpha + \gamma) + c_{z-}^{(z)}(\beta + \delta) \big)\, J_- \rho J_z - \big( c_{z+}^{(z)}(\alpha + \gamma) + c_{+z}^{(z)}(\beta + \delta) \big)\, J_z \rho J_+\\
      &- \big( c_{z-}^{(z)}(\alpha + \gamma) + c_{-z}^{(z)}(\beta + \delta) \big)\, J_z \rho J_-\Big]\Big\}+o(h^2).
\end{split}
\end{equation}
We notice that the $h$-dependent terms cannot be reorganized into a commutator. We thus have that, in order to have a physically admissible evolution, all terms which do not form a commutator have to vanish which, together with the conditions \eqref{constraint0}, leads to a standard von Neumann evolution equation
\begin{equation}
\begin{split}
    i\partial_t\rho
    &=\epsilon_0[J_z,\rho].
\end{split}
\end{equation} 
In Appendix~\ref{Jxcalculation} we show that this result holds for any choice of Hamiltonian written as a combination of Pauli matrices. We thus conclude that the choice of coefficients $\alpha=-\beta=-\gamma=\delta=\tfrac{1}{4}$ is the {\it unique linear combination that ensures a physically viable dynamics}. This choice ends up reproducing the ordinary von Neumann evolution equation, as for the specific model of $U_q(\mathfrak{su}(2))$ discussed previously. 

Let us now try to verify whether it is possible, from this generic linear combination, to obtain instead a Lindblad-type equation by considering $\alpha,\beta,\gamma,\delta\in\mathbb{C}$. Again, considering the decomposition \eqref{decomposition}, one obtains:

\begin{equation}
\begin{split}
i\partial_t\rho
&= \epsilon_0\Big\{ 
   (\alpha_0-\beta_0-\gamma_0+\delta_0)(J_z\rho-\rho J_z)
   + i(\alpha_1-\beta_1-\gamma_1+\delta_1)(J_z\rho-\rho J_z) \\
&\quad + h\Big[
   (-c_{+z}^{(z)}+c_{z+}^{(z)})\big((\alpha_0-\delta_0)\rho J_+ + (\gamma_0-\beta_0)J_+\rho\big) + (-c_{-z}^{(z)}+c_{z-}^{(z)})\big((\delta_0-\alpha_0)\rho J_- + (\beta_0-\gamma_0)J_-\rho\big) \\
&\quad - c_{++}^{(z)}(\alpha_0+\beta_0+\gamma_0+\delta_0)\,J_+\rho J_+
   - c_{--}^{(z)}(\alpha_0+\beta_0+\gamma_0+\delta_0)\,J_-\rho J_- \\
&\quad - \big(c_{+z}^{(z)}(\alpha_0+\gamma_0)+c_{z+}^{(z)}(\beta_0+\delta_0)\big)\,J_+\rho J_z - \big(c_{-z}^{(z)}(\alpha_0+\gamma_0)+c_{z-}^{(z)}(\beta_0+\delta_0)\big)\,J_-\rho J_z \\
&\quad - \big(c_{z+}^{(z)}(\alpha_0+\gamma_0)+c_{+z}^{(z)}(\beta_0+\delta_0)\big)\,J_z\rho J_+ - \big(c_{z-}^{(z)}(\alpha_0+\gamma_0)+c_{-z}^{(z)}(\beta_0+\delta_0)\big)\,J_z\rho J_-
   \Big] \\
&\quad + i h\Big[
   (-c_{+z}^{(z)}+c_{z+}^{(z)})\big((\alpha_1-\delta_1)\rho J_+ + (\gamma_1-\beta_1)J_+\rho\big) + (-c_{-z}^{(z)}+c_{z-}^{(z)})\big((\delta_1-\alpha_1)\rho J_- + (\beta_1-\gamma_1)J_-\rho\big) \\
&\quad - c_{++}^{(z)}(\alpha_1+\beta_1+\gamma_1+\delta_1)\,J_+\rho J_+
   - c_{--}^{(z)}(\alpha_1+\beta_1+\gamma_1+\delta_1)\,J_-\rho J_- \\
&\quad - \big(c_{+z}^{(z)}(\alpha_1+\gamma_1)+c_{z+}^{(z)}(\beta_1+\delta_1)\big)\,J_+\rho J_z - \big(c_{-z}^{(z)}(\alpha_1+\gamma_1)+c_{z-}^{(z)}(\beta_1+\delta_1)\big)\,J_-\rho J_z \\
&\quad - \big(c_{z+}^{(z)}(\alpha_1+\gamma_1)+c_{+z}^{(z)}(\beta_1+\delta_1)\big)\,J_z\rho J_+ - \big(c_{z-}^{(z)}(\alpha_1+\gamma_1)+c_{-z}^{(z)}(\beta_1+\delta_1)\big)\,J_z\rho J_-
   \Big]
\Big\}+o(h^2).
\end{split}
\end{equation}

On this equation we must impose the following constraints

\begin{equation}
\begin{cases}
    \alpha_1-\beta_1-\gamma_1+\delta_1=0,\\
    \alpha_1+\beta_1+\gamma_1+\delta_1=0,\\
    \alpha_1-\delta_1=\gamma_1-\beta_1=0,\\
    \alpha_1+\gamma_1=\beta_1+\delta_1=0.
\end{cases}
\end{equation}

The first condition is required in order to cancel the term $i(\alpha_1-\beta_1-\gamma_1+\delta_1)(J_z\rho-\rho J_z)$, which spoils the von Neumann form required in the $h\rightarrow 0$ limit. The second condition eliminates the terms proportional to $J_+\rho J_+$ and $J_-\rho J_-$, which cannot be recast into a valid Lindblad form $F_i\rho F_j-\frac{1}{2}\{F_j^\dagger F_i,\rho\}$. The third constraint is needed to eliminate all terms of the form $\rho J_i$ and $J_i\rho$, which by their very structure cannot be arranged into a Lindblad form. Finally, the last constraint follows from the fact that the terms $J_+\rho J_z$, $J_-\rho J_z$, $J_z\rho J_+$ and $J_z\rho J_-$ cannot be paired with the appropriate anticommutator contributions $-\tfrac12\{F_j^\dagger F_i,\rho\}$. Once all these conditions are imposed, every term arising from the imaginary part of the coefficients cancels out, and we recover exactly the same structure as in the undeformed case.

\section{Quantum evolution for $\kappa$-deformed spacetime symmetries}\label{kminkowski}

In this section we turn to the question of quantum time evolution in the context of noncommutative deformations of spacetime symmetries. As mentioned in the Introduction, it has been suggested in the literature that, in theories based on $\kappa$-deformed symmetries arising from the non-relativistic limit of the $\kappa$-Poincaré algebra—related to $\kappa$-Minkowski non-commutative spacetime \cite{Majid:1994cy}—the deformed coproducts and antipodes induce, via the adjoint action, a Lindblad-like evolution equation. We would like here to re-examine these claims in light of the analysis we presented above. We recall that the existing literature takes as a starting point the so-called {\it classical basis} of the $\kappa$-Poincaré algebra \cite{Borowiec:2009vb}. Since the symmetry group relevant for nonrelativistic quantum mechanics is the Galilei group, rather than the Poincaré group, one considers the nonrelativistic limit of the $\kappa$-Poincaré algebra, namely, the $\kappa$-Galilei algebra. For such deformed algebra, the coproducts and the antipodes for the translation generators assume the following form at leading order in the inverse deformation parameter $\kappa$
\begin{equation}\label{originalstrucutre}
\begin{split}
\Delta P_0 &= P_0\otimes\mathbbm{1}+\mathbbm{1}\otimes P_0+\frac{1}{\kappa}P_i\otimes P^i,\\
\Delta P_i &= P_i\otimes\mathbbm{1}+\mathbbm{1}\otimes P_i+\frac{1}{\kappa}P_i \otimes P_0,\\
S(P_0) &= -P_0 + \frac{1}{\kappa}\,P^2,\\
S(P_i) &= -P_i + \frac{1}{\kappa}  P_i P_0.
\end{split}
\end{equation}

Attempts to obtain Lindblad evolution equations have been made in \cite{Arzano:2022nlo}, where the authors apply the map $G\to iG$ to the symmetry generators with the intention to map the abstract elements $G$ of the $\kappa$-Galilei algebra into Hermitian operators that can be represented on the algebra of operators of a quantum system, obtaining the following coproducts and antipodes:
\begin{equation}
\begin{split}\label{coalka}
    \Delta P_0 &= P_0\otimes\mathbbm{1}+\mathbbm{1}\otimes P_0-\frac{i}{\kappa}P_i\otimes P^i,\\
    \Delta P_i &= P_i\otimes\mathbbm{1}+\mathbbm{1}\otimes P_i-\frac{i}{\kappa}P_i \otimes P_0,\\
    S(P_0) &= -P_0 - \frac{i}{\kappa}\,P^2,\\
    S(P_i) &= -P_i - \frac{i}{\kappa}  P_i P_0.
\end{split}
\end{equation}
Using the prescription \eqref{prescription1}, the coproduct and antipode for $P_0$ above lead to a Lindblad evolution. We notice here that the coalgebra structure \eqref{coalka} is not tenable as a physically viable fundamental model since it spoils the hermiticity of the coproducts of the generators, leading to complex values of total momentum and energy for a two-particle system. 

Let us now analyze the model using the standard coalgebra structure \eqref{originalstrucutre} and the general prescription \eqref{prescriptiongeneral}. Since the coproduct of $P_0$ is symmetric (at leading order in $1/\kappa$), then $\mathrm{ad}^L_{P_0}(\rho)=-\mathrm{ad}^R_{P_0}(\rho)$. This observation leads to a simplification of \eqref{prescriptiongeneral}, which becomes
\begin{equation}\label{collapsedevolution}
    i\partial_t(\rho)=(\alpha-\gamma)\mathrm{ad}^L_{P_0}(\rho)+(\beta-\delta)[\mathrm{ad}^L_{P_0}(\rho)]^\dagger.
\end{equation}
Let us consider first the case of real coefficients, in which the evolution equation collapses to the prescription \eqref{prescription1}. The left adjoint action of $P_0$ on a density matrix $\rho$ is
\begin{equation}
\begin{split}
\mathrm{ad}_{P_0}^L(\rho)&=(id\otimes S)\Delta P_0\diamond \rho\\
&= P_0\rho+\rho S(P_0)+\frac{1}{\kappa}P_i\rho S(P^i)\\
&= [P_0,\rho]+ \frac{1}{\kappa} \rho P^2- \frac{1}{\kappa} P_i \rho\, g^{ij} P_j+\mathcal{O}\!\left(\kappa^{-2}\right).
\end{split}
\end{equation}
Inserting the adjoint action into the evolution equation \eqref{collapsedevolution} we have
\begin{equation}
\begin{split}
\partial_t\rho&=-\frac{i}{2}\Big( [P_0,\rho]+ \tfrac{1}{\kappa} \rho P^2- \tfrac{1}{\kappa} P_i \rho\, g^{ij} P_j
+[P_0,\rho]- \tfrac{1}{\kappa}P^2 \rho+ \tfrac{1}{\kappa} P_i \rho\, g^{ij} P_j \Big)+\mathcal{O}\!\left(\kappa^{-2}\right)\\
&=-i[P_0,\rho] - \frac{i}{2\kappa}[\rho, P^2] +\mathcal{O}\!\left(\kappa^{-2}\right).
\end{split}
\end{equation}
Thus, at leading order, the dynamics is governed by an equation which can be written in von Neumann form
\begin{equation}\label{deformedhamiltonianevol}
    \partial_t\rho=-i[H_{\mathrm{eff}},\rho],  
\end{equation}
with
\begin{equation}\label{deformedhamiltoniankgalilei}
    H_{\mathrm{eff}}=P_0-\frac{1}{2\kappa}P^2.
\end{equation}
Thus we conclude that the prescription \eqref{1q1q1q1q} ensures a von Neumann evolution, now governed by an effective {\it deformed} Hamiltonian.

Let us finally explore the case in which the coefficients in the prescription \eqref{genericcombination} are allowed to be complex. In this case, using the definition \eqref{decomposition}, the resulting evolution is

\begin{equation}
\begin{split}
    i\partial_t\rho&=[H_{\mathrm{eff}},\rho] + i(\alpha_1-\gamma_1)\left( [P_0,\rho]+\frac{1}{\kappa}\rho P^2-\frac{1}{\kappa}P_i\rho g^{ij}P_j \right)\\
    &+i(-\beta_1+\delta_1)\left( [P_0,\rho]-\frac{1}{\kappa}P^2 \rho+\frac{1}{\kappa}P_j g^{ij}\rho P_i \right) +\mathcal{O}\!\left(\kappa^{-2}\right).
\end{split}
\end{equation}
We notice that, in order to recover the undeformed von Neumann evolution in the limit for $\kappa\to\infty$, we must impose $\alpha_1-\gamma_1=\beta_1-\delta_1=0$. Such choice leads to the same result \eqref{deformedhamiltonianevol} of the real coefficients case. We are thus led to the conclusion that the only physically viable time evolution in the case of $\kappa$-deformed symmetries is a standard von Neumannn one with deformed Hamiltonian \eqref{deformedhamiltoniankgalilei}.

\section{Summary and outlook}

In this work we investigated the possibility of obtaining deformed quantum time evolution equations for $q$-deformations of the $\mathfrak{su}(2)$ algebra and for $\kappa$-deformed spacetime symmetries. Starting from the $U_q(\mathfrak{su}(2))$ Hopf algebra, we proposed a general family of prescriptions to define the evolution of a quantum state expressed in terms of quantum adjoint actions, which in the undeformed case reduce to the standard von Neumann form. We found that only a particular combination of adjoint actions guarantees a physical evolution and leads to a standard von Neumann equation. We then constructed the most general linear Hopf algebra deformations of $su(2)$ with undeformed algebraic sector. By imposing the homomorphism and hermiticity conditions of the coproducts, we were able to simplify the structure of the coalgebra. Using the Hamiltonian $J_z$, we computed the most general linear combination of quantum adjoint actions and obtained the same result as in the $U_q(\mathfrak{su}(2))$ case, namely a standard von Neumann evolution. 

We also attempted to obtain a Lindblad-type evolution using complex coefficients in the linear combination of quantum adjoint actions, without success. Finally, we explored the possibility of obtaining a fundamental deformed quantum dynamics emerging from the $\kappa$-deformations of spacetime symmetries. Using a model of $\kappa$-Galilei algebra proposed in literature, we found that a quantum evolution leading to fundamental decoherence is not possible without spoiling the hermiticity of the coproducts of the generators. The only physically viable evolution is instead of von Neumann type governed by a deformed effective Hamiltonian.

Our results suggest that only a very specific combination of left and right adjoint actions leads to a physically viable quantum evolution when we deal with Hopf algebra deformations of $\mathfrak{su}(2)$, and that the corresponding evolution equation is of standard von Neumann type. Our analysis has been carried out on Hopf algebra models with an effective undeformed algebra sector, i.e. models in which the commutators of generators close a standard $\mathfrak{su}(2)$ Lie algebra. A natural next step is thus to extend the study to models with a deformed algebra sector, in order to test the generality of our conclusions. 

We also applied our general prescription for defining an evolution equation in models with $\kappa$-deformed spacetime symmetries. We first observed that a previously proposed prescription leading to Lindblad-like evolution in this setting leads unphysical features, and we then derived a consistent von Neumann evolution governed by a deformed Hamiltonian. A natural extension of this analysis would be to apply the same prescription to other models of noncommutative spacetimes and deformed symmetries. A particularly intriguing case is provided by the $\theta$-Minkowski noncommutative spacetime \cite{Chaichian:2004za} and its associated deformed symmetries, where the coproducts of translation generators remain undeformed while the Lorentz sector is modified. In this scenario, one could obtain a non-trivial evolution generated by boosts while retaining a standard evolution under ordinary time translations, potentially leading to novel physical effects in the comparison between quantum dynamics in uniformly accelerated and inertial frames. We leave these extensions and further investigations for future work.

\section*{Acknowledgements}
MA acknowledges support from the INFN Iniziativa Specifica QUAGRAP and from the COST Action CaLISTA CA21109. This work also falls within the scopes of the COST Action BridgeQG CA23130.

\appendix

\section{Explicit computation for $J_x$ Hamiltonian}\label{Jxcalculation}

The result shown in \ref{general adjoint action} is valid for all choices of Hamiltonians written as a combination of Pauli operators. Let us demonstrate it explicitly. We fix the following Hamiltonian
\begin{equation}
    H=\epsilon_0J_x=\epsilon_0(J_++J_-),
\end{equation}
\begin{equation}
\begin{split}
    \Delta H&=\epsilon_0\Big[J_x\otimes\mathbbm{1}+\mathbbm{1}\otimes J_x+h(c_{+z}^{(+)}J_+\otimes J_z+c_{z+}^{(+)}J_z\otimes J_+-(c_{+z}^{(z)}+c_{z+}^{(z)})J_+\otimes J_++c_{z-}^{(z)}J_+\otimes J_-\\
    &+c_{-z}^{(z)}J_-\otimes J_+-\frac{1}{4}+c_{--}^{(z)}J_-\otimes J_z-\frac{1}{4}c_{--}^{(z)}J_z\otimes J_--\frac{1}{2}(c_{+z}^{(z)}+c_{z+}^{(z)})J_z\otimes J_z-\frac{1}{4}c_{++}^{(z)}J_+\otimes J_z\\
    &-\frac{1}{4}c_{++}^{(z)}J_z\otimes J_++c_{+z}^{(z)}J_+\otimes J_-+c_{z+}^{(z)}J_-\otimes J_+-(c_{-z}^{(z)}+c_{z-}^{(z)})J_-\otimes J_--c_{z+}^{(+)}J_-\otimes J_z-c_{+z}^{(+)}J_z\otimes J_-\\
    &-\frac{1}{2}(c_{+z}^{(z)}+c_{z+}^{(z)})J_z\otimes J_z)\Big]+o(h^2).
\end{split}
\end{equation}
The left and right quantum adjoint actions are therefore
\begin{equation}
\begin{split}
    \mathrm{ad}_H^L(\rho)&=\epsilon_0\Big\{J_x\rho-\rho J_x+h[(-c_{+z}^{(+)}+c_{z+}^{(+)})\rho J_++\frac{1}{2}(c_{-z}^{(z)}-c_{z-}^{(z)})\rho J_z+(c_{z+}^{(+)}-c_{+z}^{(+)})\rho J_-+\frac{1}{2}(c_{z+}^{(z)}-c_{+z}^{(z)})\rho J_z\\
    &-c_{+z}^{(+)}J_+\rho J_z-c_{z+}^{(+)}J_z\rho J_++(c_{+z}^{(z)}+c_{z+}^{(z)})J_+\rho J_+-c_{z-}^{(z)}J_+\rho J_--c_{-z}^{(z)}J_-\rho J_++\frac{1}{4}c_{--}^{(z)}J_-\rho J_z+\frac{1}{4}c_{--}^{(z)}J_z\rho J_-\\
    &+\frac{1}{2}(c_{+z}^{(z)}+c_{z+}^{(z)})J_z\rho J_z+\frac{1}{4}c_{++}^{(z)}J_+\rho J_z+\frac{1}{4}c_{++}^{(z)}J_z\rho J_+-c_{+z}^{(z)}J_+\rho J_--c_{z+}^{(z)}J_-\rho J_++(c_{-z}^{(z)}+c_{z-}^{(z)})J_-\rho J_-\\
    &+c_{z+}^{(+)}J_-\rho J_z+c_{+z}^{(+)}J_z\rho J_-+\frac{1}{2}(c_{+z}^{(z)}+c_{z+}^{(z)})J_z\rho J_z]\Big\}+o(h^2),
\end{split}
\end{equation}

\begin{equation}
\begin{split}
    \mathrm{ad}_H^R(\rho)&=\epsilon_0\Big\{-J_x\rho+\rho J_x+h[(-c_{+z}^{(+)}+c_{z+}^{(+)})J_+\rho+\frac{1}{2}(c_{-z}^{(z)}-c_{z-}^{(z)})J_z\rho+(c_{z+}^{(+)}-c_{+z}^{(+)})J_-\rho+\frac{1}{2}(c_{z+}^{(z)}-c_{+z}^{(z)})J_z\rho\\
    &-c_{+z}^{(+)}J_+\rho J_z-c_{z+}^{(+)}J_z\rho J_++(c_{+z}^{(z)}+c_{z+}^{(z)})J_+\rho J_+-c_{z-}^{(z)}J_+\rho J_--c_{-z}^{(z)}J_-\rho J_++\frac{1}{4}c_{--}^{(z)}J_-\rho J_z+\frac{1}{4}c_{--}^{(z)}J_z\rho J_-\\
    &+\frac{1}{2}(c_{+z}^{(z)}+c_{z+}^{(z)})J_z\rho J_z+\frac{1}{4}c_{++}^{(z)}J_+\rho J_z+\frac{1}{4}c_{++}^{(z)}J_z\rho J_+-c_{+z}^{(z)}J_+\rho J_--c_{z+}^{(z)}J_-\rho J_++(c_{-z}^{(z)}+c_{z-}^{(z)})J_-\rho J_-\\
    &+c_{z+}^{(+)}J_-\rho J_z+c_{+z}^{(+)}J_z\rho J_-+\frac{1}{2}(c_{+z}^{(z)}+c_{z+}^{(z)})J_z\rho J_z]\Big\}+o(h^2).
\end{split}
\end{equation}
Computing the evolution equation according to the prescription \eqref{prescriptiongeneral}, we obtain
\begin{equation}
\begin{split}
    i\partial_t\rho&=(\alpha-\beta-\gamma+\delta)[\epsilon_0J_x,\rho]+\epsilon_0h\Big\{[(\alpha+\gamma)(-c_{+z}^{(+)}+c_{z+}^{(+)})\rho J_++(\beta+\gamma)(c_{-z}^{(-)}-c_{z-}^{(-)})J_+\rho\\
    &+(\alpha+\beta)(c_{z+}^{(z)}-c_{+z}^{(z)})\rho J_z+(\beta+\gamma)(c_{z+}^{(z)}-c_{+z}^{(z)})J_z\rho+(\alpha+\beta+\gamma+\delta)(c_{+z}^{(z)}+c_{+z}^{(z)})J_+\rho J_+\\
    &+(\alpha+\beta+\gamma+\delta)(c_{z+}^{(z)}+c_{+z}^{(z)})J_-\rho J_-+(\alpha+\beta+\gamma+\delta)(c_{+z}^{(z)}+c_{z+}^{(z)})J_z\rho J_z\\
    &+\left(-(\alpha+\gamma)c_{+z}^{(+)}+(\beta+\delta)c_{z-}^{(-)}+\frac{1}{4}(\alpha+\beta+\gamma+\delta)c_{++}^{(z)}\right)J_+\rho J_z\\
    &+\left(-(\alpha+\gamma)c_{z+}^{(+)}+(\beta+\delta)c_{-z}^{(-)}+\frac{1}{4}(\alpha+\beta+\gamma+\delta)c_{++}^{(z)}\right)J_z\rho J_+\\
    &+\left((\alpha+\gamma)c_{z+}^{(+)}-(\beta+\delta)c_{-z}^{(-)}+\frac{1}{4}(\alpha+\beta+\gamma+\delta)c_{--}^{(z)}\right)J_-\rho J_z\\
    &+\left((\alpha+\gamma)c_{+z}^{(+)}-(\beta+\delta)c_{z-}^{(-)}+\frac{1}{4}(\alpha+\beta+\gamma+\delta)c_{--}^{(z)}\right)J_z\rho J_-\\
    &-2c_{+z}^{(z)}(\alpha+\beta+\gamma+\delta)J_+\rho J_--2c_{z+}^{(-)}(\alpha+\beta+\gamma+\delta)J_-\rho J_+]\Big\}+o(h^2),
\end{split}
\end{equation}
where all deformation terms appear multiplied by the same combinations of coefficients $(\alpha+\gamma)$, $(\beta+\delta)$ and $(\alpha+\beta+\gamma+\delta)$ that were found in the $J_z$ case. It is therefore evident that, also in this case, the choice \eqref{prescriptiongeneral} ensures the standard von Neumann evolution, exactly as in the model with $J_z$ Hamiltonian. We can obtain the same result also for $H=\epsilon_0J_y$. Indeed, in the $\{J_+,J_-,J_z\}$ basis one has $J_y=\frac{1}{i}(J_+-J_-)$, so that the Hamiltonian differs from the $J_x$ case only by relative signs and phases in front of $J_\pm$. Therefore, the terms will have the same coefficients, and the resulting evolution equation has exactly the same form (standard von Neumann). It is therefore evident that, also in this case, the choice $\alpha=-\beta=-\gamma=\delta$ ensures the standard von Neumann evolution, exactly as in the model with $J_z$ Hamiltonian.\\

%It is worth recalling that, as shown in ~\cite[Sec.~2.1.2, pp.~64--65]{SakuraiNapolitano2020}, imposing unitarity on the time-evolution operator forces its generator to be proportional to a Hermitian operator. 
%When this generator acts on states in the Schrödinger picture, the resulting equation of motion takes the form of a commutator with the Hamiltonian ~\cite[Sec.~2.2.3, pp.~78--79]{SakuraiNapolitano2020}, which is the only structure that guarantees unitary evolution and preservation of positivity for all states.\\

%It is important to note, however, that an anti-Hermitian superoperator preserves the Hilbert--Schmidt norm $\mathrm{Tr}(\rho^2)$ of the state. 

 %However, introducing an imaginary deformation parameter has an immediate and serious consequence: the coproducts of the Pauli operators cease to be Hermitian \textcolor{red}{ATTENZIONE: SE ASSUMIAMO CHE LE LE SIGMA SONO NON DEFORMATE ANCHE PER h REALE I COPRDOTTI NON SONO HERMITIANI IN QUESTA BASE.}. In fact, while in the undeformed case one has $(\Delta \sigma_i)^\dagger = \Delta \sigma_i$, after the substitution $h \to i h$, this relation no longer holds. As a result, the deformed coproducts do not preserve the $*$--structure of the algebra, and operators that should represent physical observables are mapped into non-Hermitian ones. This loss of Hermiticity makes the prescription physically inconsistent. Nevertheless, in order to test whether a Lindblad-type dynamics could formally emerge, we shall analyze this scenario as well.

\bibliography{refs}    
\bibliographystyle{utphys}      

\end{document}